\documentclass[useAMS,usenatbib,twocolumn,a4paper]{mn2e}
\usepackage{xspace}
\usepackage{aas_macros}
\usepackage{url}
\usepackage{amsmath}
\usepackage{amsfonts}
\usepackage{amssymb}
\usepackage{hyperref}
\usepackage{graphicx,color}
\usepackage[capitalise]{cleveref}

\newcommand{\om}{\Omega_m}
\newcommand{\ob}{\Omega_b}
\newcommand{\sig}{\sigma_8}
\newcommand{\lcdm}{{\ensuremath{\Lambda\mathrm{CDM}}}}

\newcommand{\w}{w_0}
\newcommand{\wa}{w_a}
\renewcommand{\d}{{\rm d}}
\newcommand{\snr}{{\rm SNR}}

\newcommand{\fsky}{f_{\rm sky}}

\newcommand{\lM}{\ell_{\rm max}}

\newcommand{\jbca}{{Jodrell Bank Centre for Astrophysics, School of Physics \& Astronomy, The University of Manchester, Manchester M13 9PL, UK}}
\newcommand{\euclid}{{\it Euclid}-like}
\newcommand{\planck}{{\it Planck}}

\title[SKA Weak Lensing III: Mitigating Systematics]{SKA Weak Lensing III: Added Value of Multi-Wavelength Synergies for the Mitigation of Systematics}
\author[Camera et al.]{Stefano Camera,$^1$\thanks{E-mail: stefano.camera@manchester.ac.uk} Ian Harrison,$^1$ Anna Bonaldi$^{1,2}$ \& Michael L. Brown$^1$\\
$^1$\jbca\\
$^2$SKA Organisation, Lower Withington Macclesfield, Cheshire SK11 9DL, UK}

\begin{document}

\pagerange{\pageref{firstpage}--\pageref{lastpage}} \pubyear{2015}

\maketitle

\label{firstpage}

\begin{abstract}
In this third paper of a series on radio weak lensing for cosmology with the Square Kilometre Array, we scrutinise synergies between cosmic shear measurements in the radio and optical/near-IR bands for mitigating systematic effects. We focus on three main classes of systematics: $(i)$ experimental systematic errors in the observed shear; $(ii)$ signal contamination by intrinsic alignments;  and $(iii)$ systematic effects due to an incorrect modelling of non-linear scales. First, we show that a comprehensive, multi-wavelength analysis provides a self-calibration method for experimental systematic effects, only implying $<50\%$ increment on the errors on cosmological parameters. We also illustrate how the cross-correlation between radio and optical/near-IR surveys alone is able to remove residual systematics with variance as large as $10^{-5}$, i.e.\ the same order of magnitude of the cosmological signal. This also opens the possibility of using such a cross-correlation as a means to detect unknown experimental systematics. Secondly, we demonstrate that, thanks to polarisation information, radio weak lensing surveys will be able to mitigate contamination by intrinsic alignments, in a way similar but fully complementary to available self-calibration methods based on position-shear correlations. Lastly, we illustrate how radio weak lensing experiments, reaching higher redshifts than those accessible to optical surveys, will probe dark energy and the growth of cosmic structures in r\'egimes less contaminated by non-linearities in the matter perturbations. For instance, the higher-redshift bins of radio catalogues peak at $z\simeq0.8-1$, whereas their optical/near-IR counterparts are limited to $z\lesssim0.5-0.7$. This translates into having a cosmological signal 2 to 5 times less contaminated by non-linear perturbations.
\end{abstract}

\section{Introduction}\label{sec:introduction}
The weak gravitational lensing effect of cosmic shear is one of the main probes that the present and forthcoming generations of cosmological experiments aim at employing to perform high-precision and high-accuracy cosmology. Weak lensing is particularly valuable because it probes both the background evolution of the Universe and the growth of cosmic structures. For this reason, it has been advocated as an optimal way to test dark matter \citep{Bacon:2002hp,Taylor:2004ca,Camera:2012cj,Camera:2014rja,Shirasaki:2014noa}, dark energy \citep{Amendola:2007rr,Taylor:2006aw,Heavens:2006uk,Beynon:2011hw,Camera:2013xfa} and modified gravity \citep{Ishak:2005zs,Heavens:2007ka,Tsujikawa:2008in,Schmidt:2008hc,Beynon:2009yd,Belloso:2011ms,Camera:2009uz,Camera:2011mg,Camera:2011ms,Camera:2010wm}, in particular in combination with galaxy number counts and other observables \citep{Hu:2003pt,Jain:2007yk,Camera:2012sf,Camera:2013bwa}.

Cosmic shear surveys involve measuring correlations in the ellipticities of hundreds of thousands to tens of millions of galaxies over large areas of the sky and a wide range of redshifts \citep[see e.g.][]{Munshi:2006fn}. The main effort towards cosmic shear cosmology has hitherto focused on the optical and near-infrared (IR) bands, mainly due to the much larger number densities of background galaxies achievable at those wavelengths. Nevertheless, the possibility of weak lensing measurements in the radio band has more recently attracted increasing interest. \citet{Chang:2004ys}, who detected a cosmic shear signal in the Faint Images of the Radio Sky at Twenty cms (FIRST) survey conducted with the Very Large Array, demonstrated that weak lensing analyses can be performed with radio data. Despite the low source number density of FIRST---approximately 90 sources per square degree, 400 times smaller than in deep optical lensing surveys---a detection of cosmic shear on large scales was achieved by virtue of the large survey area covered: a quarter of the sky.

In this respect, the Square Kilometre Array (SKA) can be regarded as a game-changer, thanks to its anticipated number densities of well-detected and well-resolved galaxies of up to $\sim5$ galaxies per square arcminute over several thousand square degrees in Phase 1, and twice this number density over three quarters of the sky in Phase 2. Furthermore, the radio band offers truly unique approaches to measuring weak lensing, which are not available to optical surveys \citep[see e.g.][for an introduction]{Brown:2015ucq} and are potentially extremely powerful in minimising the most worrying systematic effects in weak lensing cosmology \citep{2016MNRAS.456.3100D}. With the aim of quantitatively assessing the potential of the SKA for weak lensing cosmology, we have embarked in a long term project, whereby the present work represents the third step.

Previously, in \citet{Harrison:2016stv} (hereafter, Paper I), we have constructed forecasts for cosmological parameter estimation. We have shown that the first phase of the SKA (SKA1) can be competitive with experiments such as the Dark Energy Survey (DES), and that the full SKA (SKA2) can potentially provide us with tighter constraints from weak lensing alone than optical cosmic shear surveys such as the Large Synoptic Survey Telescope (LSST) or the European Space Agency \textit{Euclid} satellite. Moreover, we explored the gain brought by cross-correlating shear maps between the optical and radio wavebands---a process that will be investigated further here.

Then, in \citet{Bonaldi:2016lbd} (hereafter, Paper II), we have constructed a pipeline for simulating realistic SKA weak lensing cosmology surveys. As inputs, we took: telescope sensitivity curves; correlated source flux, size and redshift distributions; a simple ionospheric model; and source redshift and ellipticity measurement errors. We have demonstrated that SKA frequency Band 2 (950$-$1760 MHz) is preferred for weak lensing science, and that an area between 1,000 and 5,000 square degrees is optimal for most SKA1 instrumental configurations, depending on observing time.

Here, we extend our analysis and scrutinise the impact of real-world effects on the deliverable science products of SKA weak lensing surveys. Indeed, so-called Stage III and IV Dark Energy Task Force (DETF; see \citealt{Albrecht:2006um}) cosmic shear experiments---of which, respectively, SKA1 and SKA2 will be representatives---will be limited not by statistical uncertainties but rather by (known and unknown) systematic effects. Therefore, we model several systematic errors that will most likely affect weak lensing surveys. We both forecast the degradation that these systematics will cause and propose ways to overcome such problems, thus recovering (most of) the cosmological information. For the sake of simplicity, we focus on one type of systematic error at a time, exploring r\'egimes where it can also be larger than the cosmological signal itself. It has to be noted, though, that when performing an actual data analysis, many different systematic effects may be present at the same time. We show that synergies between cosmic shear experiments in the radio and optical/near-IR bands are extremely effective in removing contamination from systematics.

The paper is structured as follows: in Sec.~\ref{sec:fisher}, we outline the methodology employed; in Sec~\ref{sec:sys_exp}, we focus on various types of experimental systematic errors; in Sec.~\ref{sec:sys_IAs}, we analyse contamination from intrinsic alignments (IAs); in Sec.~\ref{sec:sys_nl}, we show how radio cosmic shear surveys will be more effective in extracting cosmological information from linear scales; and in Sec.~\ref{sec:conclusions}, we summarise our results and draw our major conclusions.

\section{Methodology}
\label{sec:fisher}

\subsection{Observables}
The focus of this work is the weak lensing effect of cosmic shear (as usual, denoted by $\gamma$), whose angular power spectrum depends on the underlying cosmological model, as well as the experimental set up. Hereafter, we shall denote by $X,Y$ a survey observing at a given wavelength (radio or optical/near-IR), and indices $i,j$ will refer to redshift bins within which the redshift distribution of sources, $n_{X,Y}(z)$, have been divided into. So, $C^{X_iY_j}_\ell$ is the cross-correlation angular power spectrum of cosmic shear measured in the $i$th redshift bin of experiment $X$ and in the $j$th redshift bin of experiment $Y$. This said, we can write
\begin{equation}
C^{X_iY_j}_\ell=\frac{2\pi^2}{\ell^3}\int\!\!\d\chi\,\chi W^{X_i}(\chi)W^{Y_j}(\chi)\Delta^2_\delta\left[k_\ell(\chi),\chi\right],\label{eq:ClGG}
\end{equation}
where $\ell$ is angular scale, $\chi$ the radial comoving distance, $\Delta^2_\delta$ is the dimensionless power spectrum of density fluctuations, and $k_\ell(\chi)=\ell/\chi$ stems from Limber's approximation \citep{1953ApJ...117..134L,Kaiser:1991qi}. The $W$ functions are the shear kernels, which read
\begin{equation}
W^{X_i}(\chi)=\frac{3}{2}H_0^2\Omega_m[1+z(\chi)]\chi\int_\chi^\infty\!\!\d\chi'\frac{\chi'-\chi}{\chi'}n_{X_i}(\chi'),\label{eq:WG}
\end{equation}
with $z$ the redshift and $H_0$ the Hubble constant.

Following Paper I, as proxies for Stage III and IV DETF cosmic shear surveys we respectively adopt: the Dark Energy Survey\footnote{\url{http://darkenergysurvey.org}} (hereafter DES; \citealt{Abbott:2005bi,Abbott:2015swa}) and a \euclid\ experiment\footnote{\url{http://euclid-ec.org}} \citep{Laureijs:2011gra,Amendola:2012ys,Amendola:2016saw} in the optical/near-IR band; and SKA phase 1, and the full SKA\footnote{\url{http://skatelescope.org}} \citep{Dewdney:2009,Brown:2015ucq} at radio wavelengths. (For additional details, we refer to Sec.~3 of Paper I.)

\subsection{Fisher Matrix Analysis}
To forecast the constraining power of the various radio and optical/near-IR cosmic shear surveys we are interested in, we adopt a Fisher matrix approach. Given a likelihood function $L(\boldsymbol\vartheta)$ for a set of model parameters $\boldsymbol\vartheta=\{\vartheta_\alpha\}$, and assuming that the behaviour of the likelihood near its maximum characterises the whole likelihood function sufficiently well to be used to estimate errors on the 
model parameters \citep{Jeffreys:1961,1996ApJ...465...34V,Tegmark:1996bz}, the $1\sigma$ marginal error on parameter $\vartheta_\alpha$ reads
\begin{equation}
\sigma(\vartheta_\alpha) = \sqrt{ \left( \mathbfss F^{-1} \right)_{\alpha\alpha}},
\label{eq:marginal}
\end{equation}
where
\begin{equation}
    \mathbfss F_{\alpha\beta}=\left\langle-\frac{\partial^2\ln L(\bvartheta)}{\partial\vartheta_\alpha\partial\vartheta_\beta}\right\rangle
\end{equation}
is the Fisher matrix, $\mathbfss F^{-1}$ is its inverse, and the index `$\alpha\alpha$' denotes diagonal elements. Furthermore, the Fisher matrix approach allows us to estimate the bias, $b(\vartheta_\alpha)$, that we would get on a parameter's best-fit value if we neglected some other (e.g.\ systematic) parameter in the analysis. The calculation of such a bias is presented and discussed in Appendix~\ref{sec:Fisher_bias}.

Our data will come from the measurement of the (auto- or cross-) correlation angular power spectra $C^{X_iY_j}_\ell$ between the observed shear inferred from galaxy ellipticities for radio and optical/near-IR experiments $X$ and $Y$ in the redshift bins $i$ and $j$ (see Paper I for details). In the presence of noise, $\mathcal N^{X_iY_j}_\ell$, the observed cosmic shear power spectrum is
\begin{equation}
\widehat{C}^{X_iY_j}_\ell=C^{X_iY_j}_\ell+\mathcal N^{X_iY_j}_\ell.
\end{equation}
Then, to translate the Fisher matrix to the space of the model parameters, $\boldsymbol\vartheta$, it is sufficient to multiply the inverse of the covariance matrix by the Jacobian of the change of variables, viz.\
\begin{equation}
\mathbfss{F}_{\alpha\beta} = \sum_{\ell,\ell'=\ell_\mathrm{min}}^{\ell_\mathrm{max}}
\frac{\partial \mathbfss{C}^{XY}_\ell}{\partial \vartheta_\alpha}
\left[ \mathbf{\Gamma}^{XY}_{\ell\ell'} \right]^{-1}
\frac{\partial \mathbfss{C}^{XY}_{\ell'}}{\partial \vartheta_\beta},
\label{eq:fisher}
\end{equation}
where $\mathbf{\Gamma}^{XY}_{\ell\ell'}$ is the data covariance (assumed to be diagonal in $\ell-\ell'$). As in Paper I, we consider the standard \lcdm\ parameter vector $\boldsymbol\vartheta=\{\om,\ob,h,n_s,\sig\}$, to which we append the dark energy equation-of-state parameters, $\{w_0,w_a\}$, when we quote dark energy forecasts. In the following, we fix $\ell_{\rm min}=20$ and $\ell_{\rm max}=3000$.

For computational simplicity, the $\ell$-diagonal matrix $\mathbf{\Gamma}^{XY}_{\ell\ell'}$ only represents the Gaussian part of the total covariance matrix, whose other terms are a non-Gaussian part, coming form the trispectrum, and the so-called super-sample variance. Employing the full covariance matrix is not fully equivalent to the simplified Gaussian case usually adopted here and in Fisher matrix analyses. However, we emphasise that we do not extend our analysis to the strongly non-linear r\'egime of perturbations and that, at the $\lM$ considered here, marginal errors forecast with the Gaussian and non-Gaussian covariance are still in good agreement \citep[e.g.][]{Kiessling:2011gv}.

Given the large number of spectra necessary for all the experiments and cross-correlations we employ, as well as the many (cosmological and nuisance) parameters investigated, we have extensively tested the stability and reliability of our Fisher matrices in various ways. A comprehensive description of our method is outlined in Appendix~\ref{sec:Fisher_stability}. Moreover, we cross-checked our Fisher matrix procedure against the Monte Carlo Markov Chains simulations used in Paper I. The detailed results of this comparison is presented in Appendix~\ref{sec:Fisher_comparison}, and the general agreement is very good: the scatter between the two methods is generally smaller than $10\%$ but for the parameters with the most non-Gaussian contours and some configurations where the r\^ole of priors is particularly important.

\section{Experimental Systematics}\label{sec:sys_exp}
As emphasised in \citet{Brown:2015ucq}, radio and optical weak lensing surveys have a particularly useful synergy in quantifying and reducing the impact of systematic effects that may dominate each survey alone. Here, we explore to what extent the cross-correlation of the shear estimators from one of these surveys with those of the other will mitigate the impact of several systematic errors.

Starting from the complex shear $\gamma=\gamma_1+i\gamma_2$ at a given 3D position on the sky, $(\theta,z)$, we assume that the measured shear contains the cosmological signal $\gamma$ plus a systematic error, viz. $\gamma^{\rm obs}=\gamma+\gamma^{\rm sys}$. Here, we assume that the shear systematic error can be decomposed into residual systematics and a calibration error (often called additive and a multiplicative terms; cfr \citealt{Heymans:2005rv,Huterer:2005ez,Massey:2006ha,Amara:2007as}), namely
\begin{equation}
\gamma^{\rm sys}(\theta,z)=\gamma^{\rm mul}(z)\gamma(\theta,z)+\gamma^{\rm add}(\theta,z).\label{eq:shear_sys1}
\end{equation}
Under the assumption of no noise and of small multiplicative systematics, this leads to an observed power spectrum of the form
\begin{multline}
C^{\rm obs}_\ell(z,z')=\\\left\{1+\left[\gamma^{\rm mul}(z)+\gamma^{\rm mul}(z')\right]\right\}C_\ell(z,z')+C^{\rm add}_\ell(z,z').\label{eq:Clobswsys}
\end{multline}

Whilst a full treatment of the various effects that could lead to experimental systematics is strongly survey-dependent and is beyond the purpose of this paper \citep[see e.g.][for a comprehensive analysis of optical survey systematics]{Cardone:2013lha}, we shall here focus on some general shapes of the systematics power spectrum that is possibly more degenerate with the cosmological signal we are after. The purpose of this is to quantify the amelioration brought by the cross-correlation of radio and optical/near-IR cosmic shear measurements.

\subsection{Residual Systematics}\label{sec:sys_add}
Our approach is to be agnostic about the origin of the systematic effects, and instead parameterise them with an $\ell$-dependence. Following \citet{Amara:2007as}, we define
\begin{equation}
C^{\rm add}_\ell=A_{\rm add}\frac{n_{\rm add}\log(\ell/\ell_{\rm add})+1}{\ell(\ell+1)},\label{eq:Clsys_add1}
\end{equation}
which allows for the possibility that the residual systematics power spectrum can be positive or negative, and that it may transit from one to the other. More specifically, $\ell(\ell+1)C^{\rm add}_\ell$, whose slope is $n_{\rm add}A_{\rm add}$, scales linearly with $\log\ell$ and amounts to $A_{\rm add}$ at $\ell_{\rm add}$. For practical purposes, we set the amplitude $A_{\rm add}$---the additional parameter that we marginalise over---according to the systematics signal variance
\begin{equation}
\sigma_{\rm sys}^2=\int\!\!\frac{\d\ln\ell}{2\pi}\,\ell(\ell+1)\left|C^{\rm add}_\ell\right|.\label{eq:sys2}
\end{equation}

\begin{figure}
\centering
\includegraphics[width=\columnwidth]{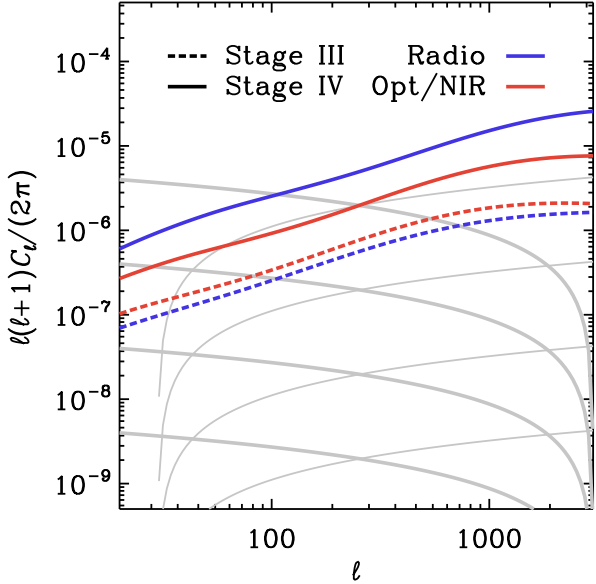}
\caption{Residual (additive) systematics power spectra (grey curves) for $\sigma_{\rm sys}^2=10^{-8}$, $10^{-7}$, $10^{-6}$ and $10^{-5}$ (from bottom to top), $\ell_{\rm add}=300$ and $n_{\rm add}=\pm1$ (thin/thick lines). As a comparison, the cosmic shear angular power spectra with the smallest power (viz.\ for the lowest-redshift bins) are shown for Stage III (dashed) and IV (solid) DETF experiments (blue for radio and red for optical/near-IR surveys).}\label{fig:Cl_add1}
\end{figure}
To have a better grasp of how such residual systematic effects will look at the level of the power spectrum, Fig.~\ref{fig:Cl_add1} depicts Eq.~\eqref{eq:Clsys_add1}, where the grey curves are $C^{\rm add}_\ell$ for $\sigma_{\rm sys}^2=10^{-8}$, $10^{-7}$, $10^{-6}$ and $10^{-5}$ from bottom to top. We choose $\ell_{\rm add}=300$, and thin/thick lines are for $n_{\rm add}=\pm1$. As a comparison, the blue and red curves respectively show the shear power spectrum for radio and optical/near-IR alone in the the first redshift bin auto-correlation. Note that, since weak lensing is an integrated signal along the line of sight---i.e.\ it increases with redshift---the cosmic shear power spectra we depict here are those with the smallest signal, namely those for which the contamination from the systematics power spectrum is the largest.

\begin{figure*}
    \centering
    \includegraphics[width=\textwidth]{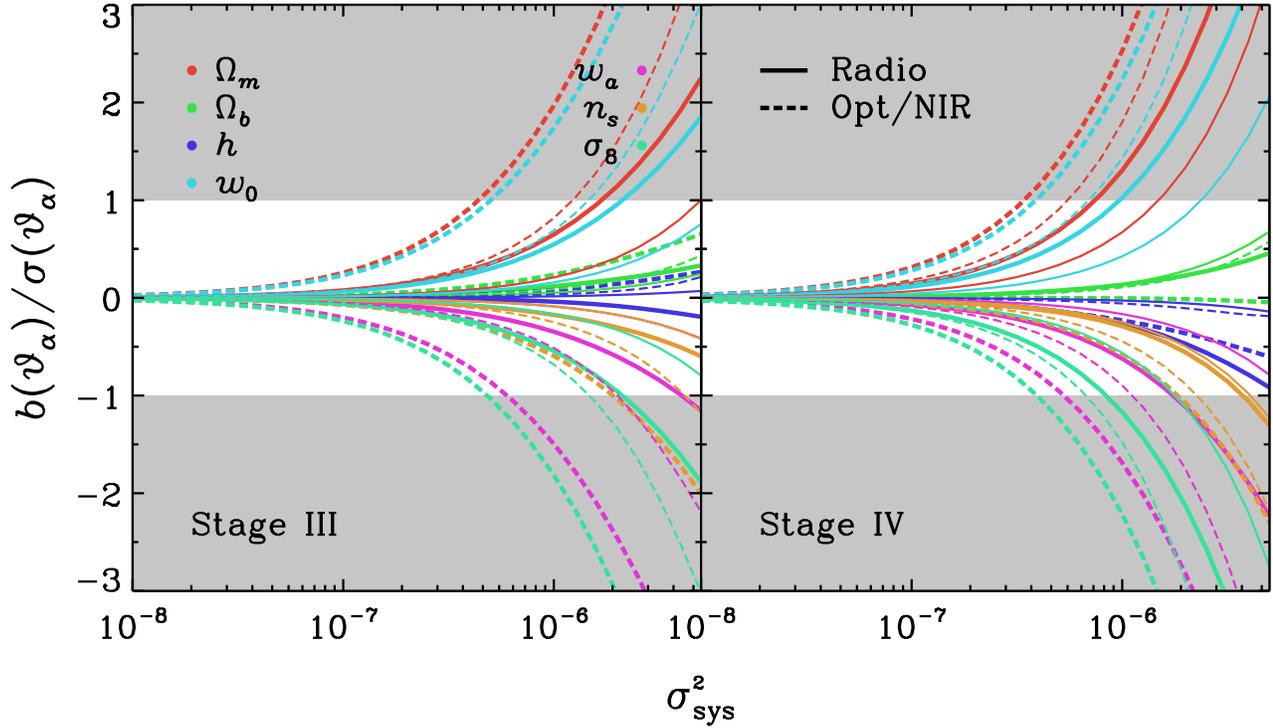}
    \caption{Normalised bias on cosmological parameters, $b(\vartheta_\alpha)/\sigma(\vartheta_\alpha)$, as a function of the systematic signal variance, $\sigma_{\rm sys}^2$. The left(right) panel is for Stage III(IV) DETF experiments, with solid(dashed) curves for optical/near-IR(radio) cosmic shear measurements. Each colour refers to a specific cosmological parameter, whilst the thickness of the curve is related to the slope of systematics power spectrum (see text). The central, horizontal white band denotes a 1$\sigma$ confidence interval within which the statistical error dominates over the bias.}
    \label{fig:sys_add1}
\end{figure*}

Figure~\ref{fig:sys_add1} shows the impact of this type of additive systematic effect. The bias $b(\vartheta_\alpha)$ on the reconstruction of each \lcdm\ and dark energy cosmological parameter in units of the forecast precision on the measurement, $\sigma(\vartheta_\alpha)$, is shown as a function of the systematics signal variance defined in Eq.~\eqref{eq:sys2}, for Stage III and IV DETF cosmic shear surveys (left and right panel, respectively). The horizontal white band represents a `safety' area where such additive systematics lead to a bias in cosmological parameter reconstruction within 1$\sigma$ of its true value. In other words, this sets a requirement on optical/near-IR and radio cosmic shear surveys in order not to severely bias cosmological parameter reconstruction. As above, $\ell_{\rm add}=300$, and we have explored a wide range of $n_{\rm add}$ values as in Fig.~\ref{fig:Cl_add1}, from steep and negative slopes ($n_{\rm add}=-1$, thick lines), to positive tilts ($n_{\rm add}=1$, thin lines). We find that the magnitude of the effect and the impact on parameter reconstruction, albeit at a first glance rather constant, has a non-negligible increment for Stage IV surveys. This is due to the higher sensitivity of such next-generation experiments to cosmological parameters---that is to say, their smaller $\sigma(\vartheta_\alpha)$.

\begin{figure*}
    \centering
    \includegraphics[width=\textwidth]{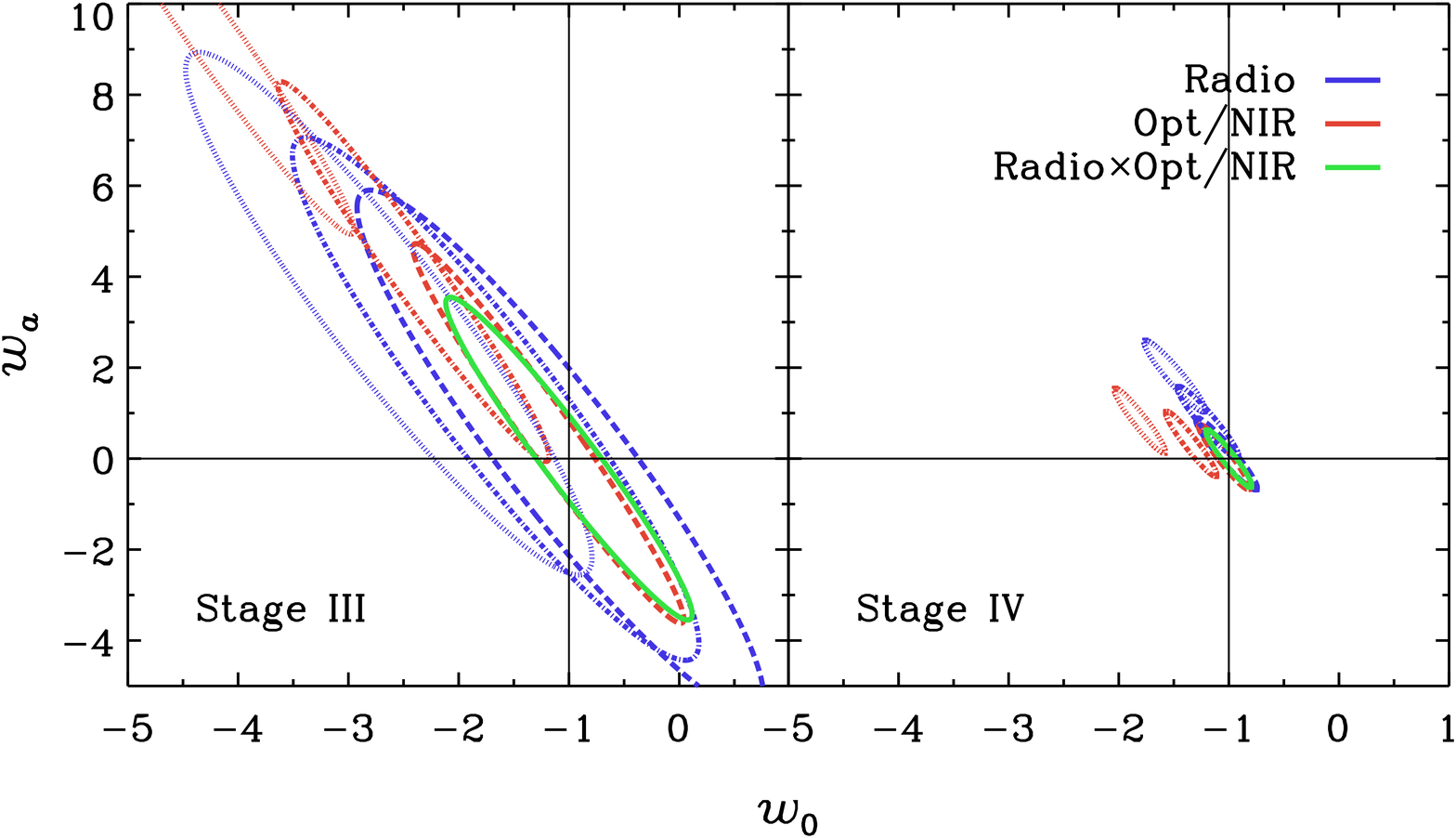}
    \caption{Marginal joint 1$\sigma$ error contours in the dark energy equation-of-state parameter plane. The black cross indicates the \lcdm\ fiducial values for dark energy parameters, namely $\{w_0,w_a\}=\{-1,0\}$. Blue, red and green ellipses are for radio and optical/near-IR surveys and their cross-correlation, respectively. The left(right) panel is for Stage III(IV) DETF cosmic shear surveys. Dashed, dot-dashed and dotted contours refer to amplitudes of the residual systematic power spectrum with variance $\sigma_{\rm sys}^2=10^{-7}$, $10^{-6}$ and $5\times10^{-5}$, respectively. All contours but those for the cross-correlation are biased (i.e.\ they are not centred on the black cross) due to the presence of residual, additive experimental systematics (Sec.~\ref{sec:sys_add}).}
    \label{fig:sys_add}
\end{figure*}
Figure~\ref{fig:sys_add} shows the impact of this residual systematics to the reconstruction of dark energy equation-of-state parameters, $\{w_0,w_a\}$. Left(right) panels refer to Stage III(IV) DETF cosmic shear experiments, with radio, optical/near-IR and their cross-correlation respectively in blue, red and green. The black cross indicates the \lcdm\ fiducial value, i.e.\ $\{w_0,w_a\}=\{-1,0\}$. The green solid, correctly centred ellipse shows the unbiased result obtained with the cross-correlation of optical/near-IR and radio surveys (insensitive to additive systematics), whereas dashed, dot-dashed and dotted ellipses are for the residual systematic power spectrum with $n_{\rm add}=-1$ and variance $\sigma_{\rm sys}^2=10^{-7}$, $10^{-6}$ and $5\times10^{-5}$, respectively. It appears to be a worrying scenario, particularly for Stage IV DETF surveys (right panel). This is mainly because such experiments will reach an exquisite accuracy in the measurement of the various cosmological parameters, being therefore more prone to even small biases due to systematics. In this case, the added value of the cross-correlation cosmic shear power spectrum of radio and optical/near-IR is straightforward to appreciate.

\subsection{Calibration errors}\label{sec:sys_mul}
As also discussed by \citet{Huterer:2005ez}, a calibration systematic error in the estimate of the shear $\gamma$ leads to a multiplicative spurious term represented by the term in square brackets in Eq.~\eqref{eq:Clobswsys}. Such a calibration error manifests itself as an overall amplitude factorising the cosmological signal, viz.\ $\gamma^{\rm mul}(z)\gamma(\theta,z)$. There are, however, two important differences compared to the previous case of residual (or additive) systematics. First, a calibration error term will be also present in the cross-correlation power spectrum. This is because this multiplicative systematic term, being attached to the cosmological signal in the fashion of an overall amplitude, will not cancel out when correlating data sets obtained in different bands of the electromagnetic spectrum---opposite to what will happen for the residual (additive) systematic effect discussed in Sec.~\ref{sec:sys_add}. Secondly, such a term will most likely present a redshift-bin dependence, inherited from $\gamma^{\rm mul}(z)$. Nevertheless, it is important to emphasise that the multiplicative calibration error $\gamma^{\rm mul}(z)$ will be different for radio and optical/near-IR, and the cross-correlation of the measurements will bear a combination of the two. Therefore, in the worse-case scenario where the calibration error is so severe as to seriously threaten the precision of parameter estimation, the confidence regions for radio or optical/near-IR auto-correlations (shown for instance in Fig.~\ref{fig:sys_add}) will be scattered around the parameter space with no apparent correlation, whereas the cross-correlation of the two will contain information on both calibration errors. Hence, an \textit{a posteriori} reconstruction can be performed, where we could iteratively try to remove two multiplicative systematic effects, i.e.\ for radio and optical/near-IR data, by using three variables, namely the two auto-correlation cosmic shear power spectra and their cross-correlation.

\begin{figure}
    \centering
    \includegraphics[width=\columnwidth]{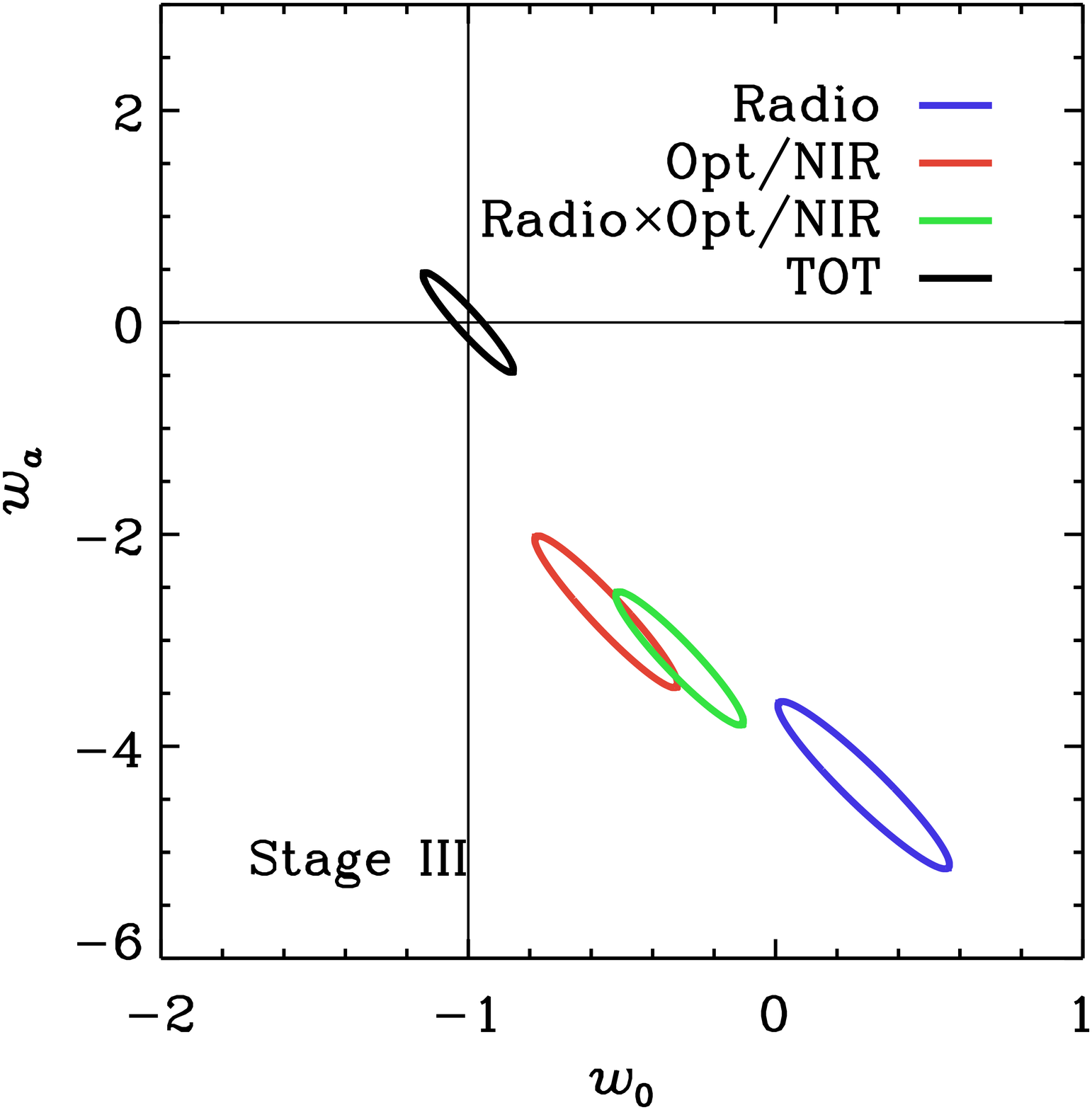}
    \caption{Same as the right panels of Fig.~\ref{fig:sys_add}, but for calibration errors (Sec.~\ref{sec:sys_mul}). Note that, in this case, the contours obtained via the cross-correlation of DES and SKA1 too is biased. Conversely, the self-calibrated combination of all auto- and cross-correlations, with the inclusion of nuisance parameters for calibration errors, is not (black ellipse).}
    \label{fig:sys_mul}
\end{figure}

To illustrate this, we generate 20 random calibration errors $\gamma_{X,i}^{\rm mul}$, 10 for the 10 radio redshift bins and 10 for the 10 optical/near-IR bins, (uniformly) randomly picked in the range $[0\%,10\%]$. By doing so, we construct a matrix $\mathbfss M$, with entries
\begin{equation}
    M_{ij}=A_{\rm mul}\left(\gamma_{X_i}^{\rm mul}+\gamma_{Y_j}^{\rm mul}\right),
\end{equation}
and overall amplitude parameter $A_{\rm mul}$, which we marginalise over. This matrix multiplies the cosmic shear tomographic matrix $\mathbfss C^{XY}_\ell$. The results are presented in Fig.~\ref{fig:sys_mul}, where, as opposed to Fig.~\ref{fig:sys_add}, the green ellipse of the cross-correlation of radio and optical/near-IR surveys is biased as well as those of the two auto-correlations. To overcome this issue we can implement the \textit{a posteriori} reconstruction discussed above. To do so, we put all the information together. In other words, we perform the Fisher analysis for a single data vector
\begin{equation}
\left(\begin{array}{c}
     \mathbfss C^{XX}_\ell \\
     \mathbfss C^{XY}_\ell \\
     \mathbfss C^{YY}_\ell
\end{array}\right),
\end{equation}
whose covariance, modulo a factor of $\delta^K_{\ell\ell'}/[(2\ell+1)\fsky]$, reads
\begin{equation}
\left(\begin{array}{ccc}
     2\left(\widehat{\mathbfss C}^{XX}_\ell\right)^2 & 2\widehat{\mathbfss C}^{XX}_\ell\widehat{\mathbfss C}^{XY}_\ell & 2\left(\widehat{\mathbfss C}^{YY}_\ell\right)^2 \\
     
     2\widehat{\mathbfss C}^{XX}_\ell\widehat{\mathbfss C}^{XY}_\ell & \left(\widehat{\mathbfss C}^{XY}_\ell\right)^2+\widehat{\mathbfss C}^{XX}_\ell\widehat{\mathbfss C}^{YY}_\ell & 2\widehat{\mathbfss C}^{XY}_\ell\widehat{\mathbfss C}^{YY}_\ell \\
     
     2\left(\widehat{\mathbfss C}^{XY}_\ell\right)^2 & 2\widehat{\mathbfss C}^{XY}_\ell\widehat{\mathbfss C}^{YY}_\ell & 2\left(\widehat{\mathbfss C}^{YY}_\ell\right)^2
\end{array}\right).
\end{equation}
The black ellipse shows the result for such a combination of auto- and cross-correlations. It is worth noting that, despite including one additional nuisance parameter, $A_{\rm mul}$, the black contour is tighter than the single blue, red and green ellipses, thanks to the fact that it encodes all the available information.

We can also be more conservative and, instead of $A_{\rm mul}$, consider 20 nuisance parameters representing the amplitudes of the 10 calibration errors for optical/near-IR measurements and the 10 ones for radio shear estimates, then marginalising over them all. Although the outcome of such a more conservative analysis is obviously less constraining than the previous case, we find than forecast errors on cosmological parameters only increase by less than $50\%$ with respect to the case with no nuisance parameters. This is a remarkable result, showing the usefulness of this self-calibration method, for which, we emphasise, the result is not biased due to any systematic effects.
%

\section{Intrinsic Alignments}\label{sec:sys_IAs}
So far, we have considered that the only spurious term in a measurement of cosmic shear is due to an experimental systematic error. However, another, well-known source of systematics is due to IAs of galaxy orientations. Indeed, the shapes of galaxies are not truly randomly oriented on the sky. During the processes leading to the formation of galaxies and their evolution, environmental effects such as tidal gravitational fields in the large-scale structure tend to align nearby galaxies. Furthermore, events such as galaxy mergers affect the relative alignments of both galaxies' shapes and angular momenta throughout their history \citep[see][for a recent review series]{Joachimi:2015mma,Kiessling:2015sma,Kirk:2015nma}. If we now focus on IAs only, the observed galaxy ellipticity is $\epsilon^{\rm obs}=\gamma+\epsilon^{\rm int}$, where $\epsilon^{\rm int}$ is the intrinsic ellipticity of a given galaxy. Therefore, when we correlate observed ellipticities, we obtain
\begin{equation}
\langle\epsilon^{\rm obs}\epsilon^{\rm obs}\rangle=\langle\gamma\gamma\rangle+2\langle\gamma\epsilon^{\rm int}\rangle+\langle\epsilon^{\rm int}\epsilon^{\rm int}\rangle.
\end{equation}
The first term is the cosmological weak lensing shear we are after, whilst the second and third terms are the very contaminants which we refer to when we talk of IAs. Usually, they are called `GI' and `II' terms, since they are correlations between the gravitational lensing signal (G) and the intrinsic shape (I).

Given the cosmic shear term (GG) defined in Eq.~\eqref{eq:ClGG}, the projected angular power spectra for the IA terms are
\begin{align}
    C^{X_iY_j}_{\mathrm{(GI)}\ell}&=\frac{2\pi^2}{\ell^3}\int\!\!\d\chi\,\chi W^{X_i}(\chi)n_{Y_j}(\chi)\Delta^2_{\rm GI}\left[k_\ell(\chi),\chi\right],\label{eq:ClGI}\\
    C^{X_iY_j}_{\mathrm{(II)}\ell}&=\frac{2\pi^2}{\ell^3}\int\!\!\d\chi\,\chi n_{X_i}(\chi)n_{Y_j}(\chi)\Delta^2_{\rm II}\left[k_\ell(\chi),\chi\right].\label{eq:ClII}\\
\end{align}
Basically, in the GI spectrum one of the lensing kernels is replaced by the galaxy redshift distribution of the sources, whereas both of them are replaced in the II spectrum. The main unknown in these expressions are the IA power spectra, $\Delta^2_{\rm GI}\left[k_\ell(\chi),\chi\right]$ and $\Delta^2_{\rm II}\left[k_\ell(\chi),\chi\right]$. As a reference, we here adopt the non-linear IA model often dubbed `corrected \citeauthor{Bridle:2007ft}' \citep[see also][]{Hirata:2004gc,Bridle:2007ft,Kirk:2011aw,Blazek:2015lfa}, where they read
\begin{align}
    \Delta^2_{\rm GI}(k,\chi)&=-C_1\frac{\bar\rho(\chi)}{D(\chi)}\Delta^2_\delta(k,\chi),\\
    \Delta^2_{\rm II}(k,\chi)&=\left[-C_1\frac{\bar\rho(\chi)}{D(\chi)}\right]^2\Delta^2_\delta(k,\chi),
\end{align}
with $\bar\rho[\chi(z)]$ the background energy density at redshift $z$, and $D$ the linear growth factor. Here, $C_1$ is the normalisation of the IA contribution, for which we use a fiducial value of $5\times10^{-14}h^{-2}\,\mathrm{Mpc}^3/M_\odot$, following \citet{Bridle:2007ft} who matched the power spectra based on the measurement of the II signal by \citet{Brown:2000gt}.

\begin{figure*}
    \centering
    \includegraphics[height=0.3\textheight]{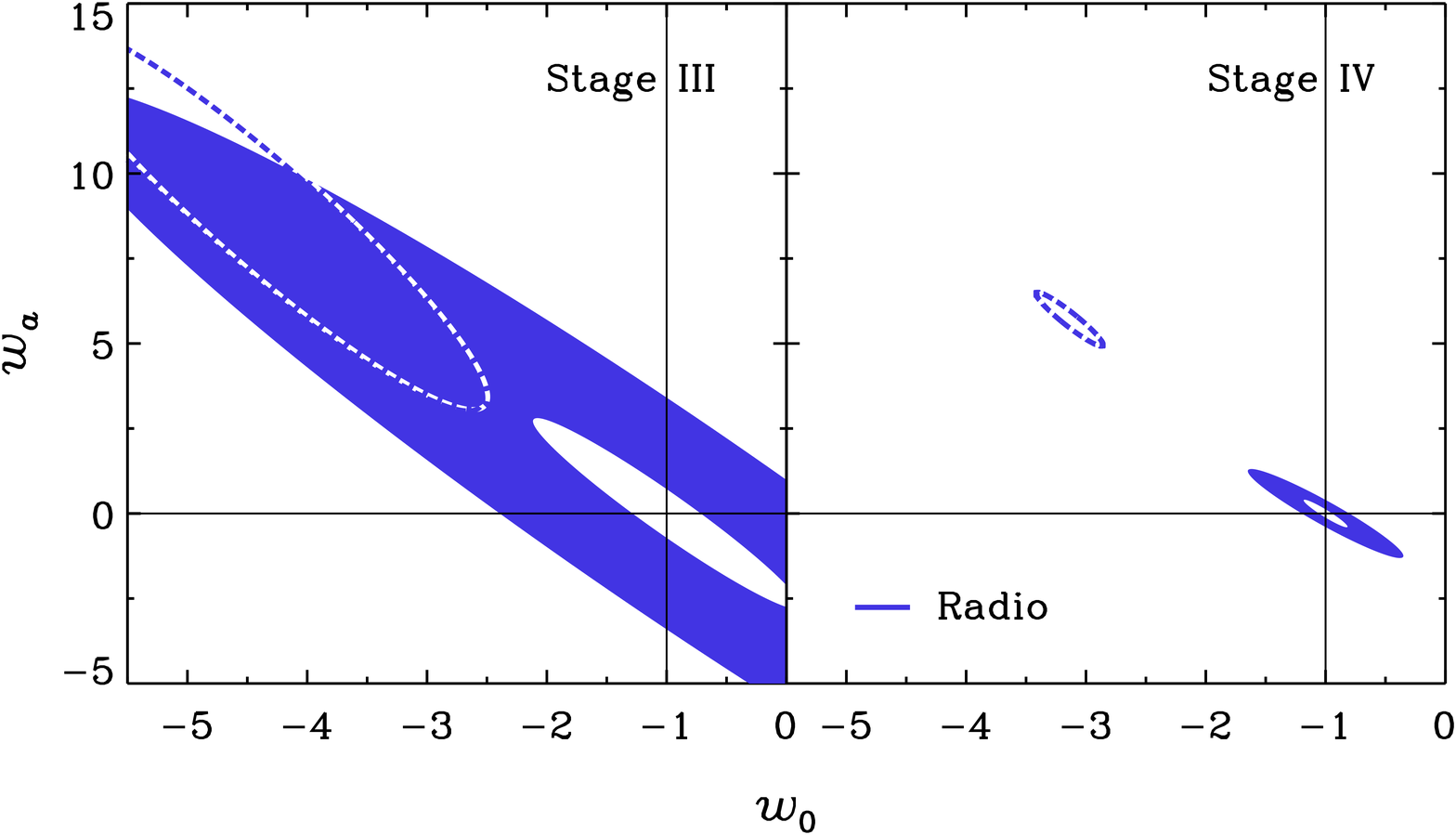}\\
    \includegraphics[height=0.3\textheight]{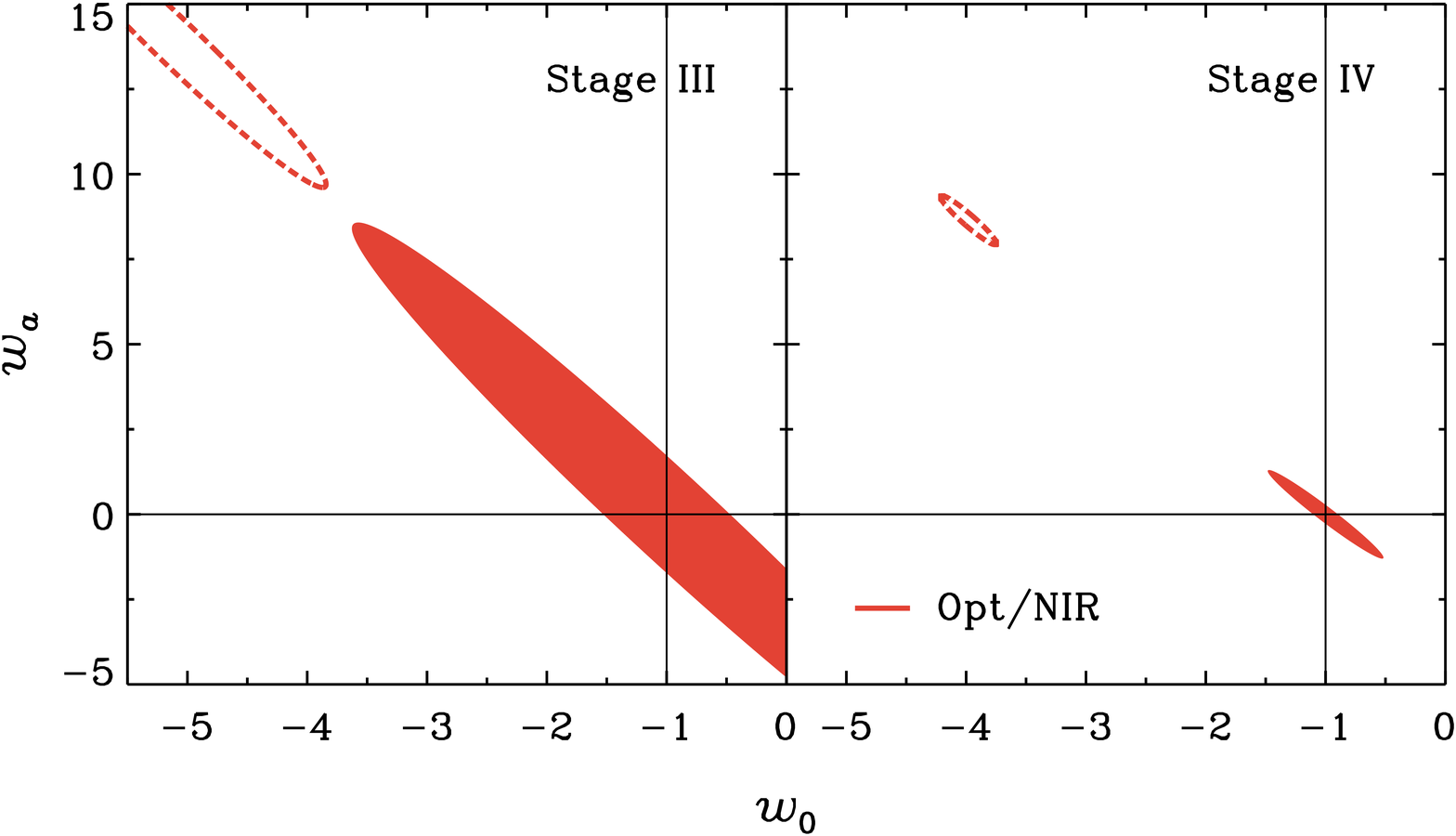}\\
    \includegraphics[height=0.3\textheight]{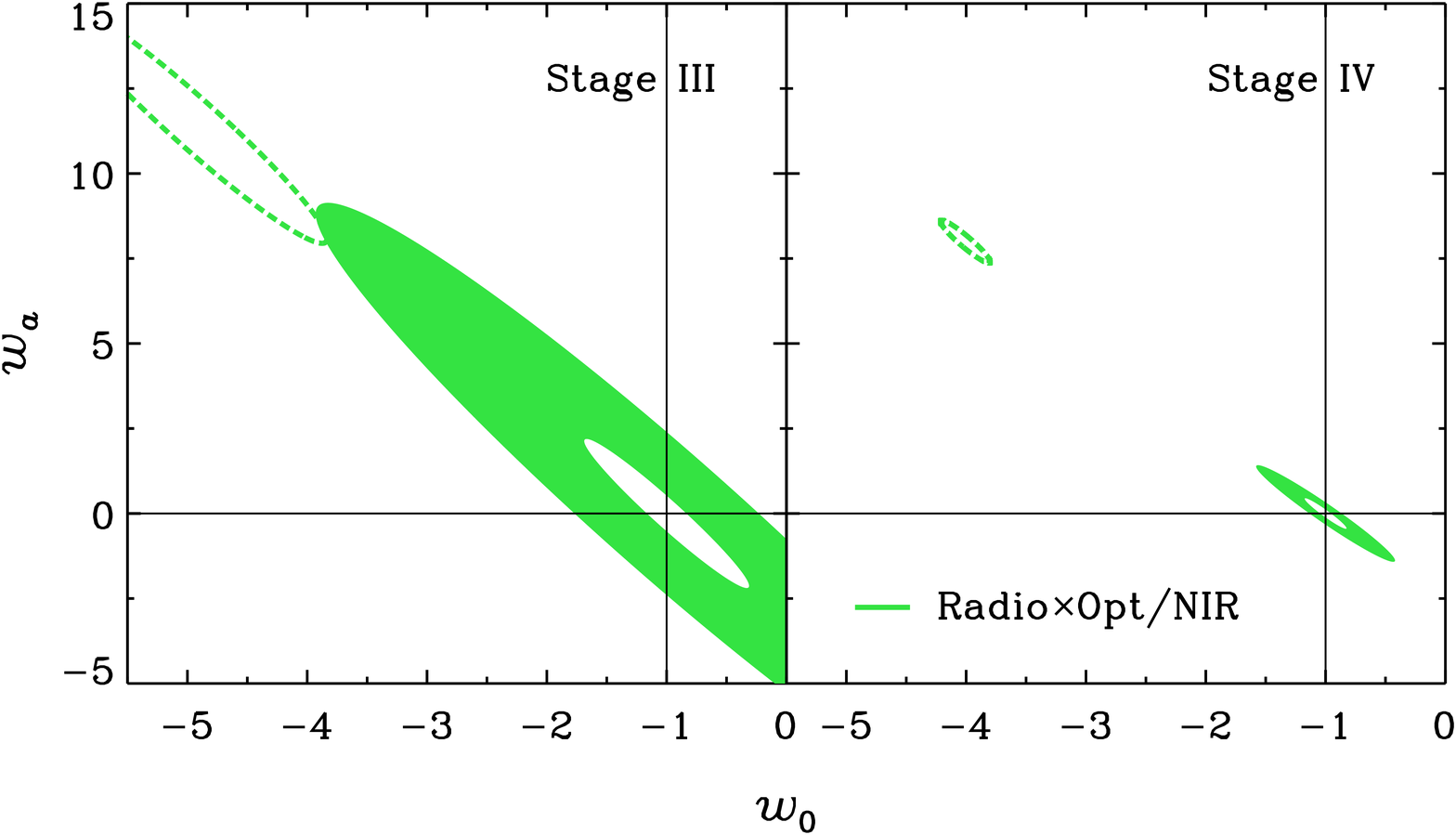}
    \caption{Similar to Fig.~\ref{fig:sys_add}, but with the systematic effect causing the bias being IAs. Plots from top to bottom show results for radio (blue) and optica/near-IR (red) surveys as well as their cross-correlation (green). Here, filled, coloured contours are for the de-biased case in which we add 40 IA nuisance parameter, whereas the innermost, empty contours refer to a case in which $95\%$ of radio sources have polarisation information and $5^\circ$ intrinsic orientation scatter.}
    \label{fig:ellipses_IAs}
\end{figure*}

Figure~\ref{fig:ellipses_IAs} illustrates the impact of neglecting IAs in the reconstruction of the dark energy equation-of-state parameters $\{w_0,w_a\}$. As in Fig.~\ref{fig:sys_add}, left(right) panels are for Stage III(IV) DETF cosmic shear experiments, whereas red, blue and green respectively refer to optical/near-IR and radio surveys and their cross-correlation. The cross indicates the \lcdm\ fiducial value of $\{w_0,w_a\}=\{-1,0\}$. Dashed contours show the best-fit confidence region that will be erroneously reconstructed if IAs were neglected in the analysis, whereas filled, coloured contours (correctly centred at values of a cosmological constant) are for the case where we introduce IA nuisance parameters and marginalise over them. Specifically, we here consider two types of nuisance parameters: $(i)$ a bias $b_I$ related to the power spectrum of the II term with respect to the matter power spectrum; and $(ii)$ a correlation coefficient $r_I$ related to the cross-correlation GI terms. Basically, each II power spectrum, for all the redshift bin combinations $i-j$, brings a $b_I^ib_I^j$ factor, whereas GI power spectra are multiplied by $b_I^ir_I^j$. This means that we add 40 nuisance parameters to our \lcdm\ plus dark energy 7-parameter set. Such a number of nuisance parameters is the reason for the broadening of the solid ellipses with respect to the dashed contours.

Here, our focus is the effect of neglecting IAs and ways by which multi-wavelength synergies can help mitigating this issue. However, it is worth to note that the true functional form of the GI and II IA power spectra is unknown. By including bias and correlation coefficients in the analysis we take into account an unknown amplitude for the IA signal, but we implicitly assume that the shape of the power spectrum is known \textit{a priori}. To appreciate the impact of different IA models, we refer the reader to e.g.\ \citet{Kirk:2011aw} and \citet{Krause:2015jqa}, who also find biases comparable to ours.

\subsection{Used of Polarisation Information}
Figure~\ref{fig:ellipses_IAs} also shows the amelioration brought by cosmic shear measurements in the radio band due to the use of source polarisation information. Following \citet{Brown:2010rr}, we can define a new estimator for the cosmic shear signal based on a galaxy's polarisation position angle $\alpha$. This quantity can indeed be measured if polarisation information is available, thanks to the fact that scalar perturbations such as density inhomogeneities distort background galaxy images but not their polarisation orientation. Hence, if the intrinsic scatter in a measurement of the polarisation position angle is $\alpha_{\rm rms}$ and a radio cosmic shear catalogue has polarisation information for a fraction $f_{\rm pol}$ of the total number of galaxies for which it has shear estimates, such a new estimator has a noise
\begin{equation}
\mathcal N_{\rm pol}^{X_iY_j}=\delta^K_{ij}\frac{\left(4\alpha_{\rm rms}\sigma_\epsilon\right)^2}{f_{\rm pol}N^i_g}.\label{eq:noise_pol}
\end{equation}
This relates to the usual noise in a cosmic shear measurement as $\mathcal N^{X_iY_j}=f_{\rm pol}/(4\alpha_{\rm rms})^2\times\mathcal N_{\rm pol}^{X_iY_j}$. The major advantage of this when dealing with IAs is the fact that, not being gravitationally lensed, a galaxy's position angle is an unbiased estimator of its true orientation in the sky. This means that, for those galaxies for which we have polarisation information, their cosmic shear power spectrum, recovered via the \citet{Brown:2010rr} estimator, does not contain IA contributions. We can therefore use it to remove contamination from IAs. In Fig.~\ref{fig:ellipses_IAs}, alongside the brute-force approach where we add nuisance parameters to account for IAs (filled, coloured contours), we also present an idealistic case in which $95\%$ of radio sources have polarisation information and their intrinsic orientation scatter is $\alpha_{\rm rms}=5^\circ$ (empty contours). Clearly, the cosmological information is recovered to high accuracy. Even better, a comparison between the solid, thick ellipses and the solid, thin ones shows that the former is even tighter than the latter. This is because, whereas the latter does not take IAs into account at all, the former takes advantage of the cosmological information encoded in the II and GI power spectra. Indeed, being proportional to the matter power spectrum, they too contain valuable cosmological information.

Clearly, the thin, solid ellipses in the middle and bottom rows of Fig.~\ref{fig:ellipses_IAs} refer to a very optimistic case. Nonetheless, this gives us an idea of the amount of information that could in principle be recovered thanks to the polarisation information in radio cosmic shear measurements. In a sense, the area within the outer and the innermost ellipse is a proxy of all the various, more realistic cases between no polarisation information at all and the idealistic scenario represented by the thin contours. In the following section, we shall outline the state of the art in this respect. As a final remark, we emphasise that the use of polarisation information is wholly complementary to more standard techniques such as so-called `self-calibration', where the clustering information on galaxies' three-dimensional position is exploited to put constraints on the IA nuisance parameters \citep[e.g.][]{Kirk:2011aw}.

\subsection{Prospects for Polarisation Measurements}
The polarisation properties of star-forming galaxies are still poorly known. The polarised signal is typically only a few percent of the total brightness, which means that very deep observations (down to the $\mu$Jy level) are needed to collect large samples. Based on a three-dimensional model of the Milky Way, \citet{2012A&A...543A.127S} predicts integrated polarisation fractions of $\sim$4.2\,\% at 4.8\,GHz and $\sim$0.8\,\% at 1.4\,GHz. The former value is consistent with the analysis of local spiral galaxies at 4.8\,GHz by \cite{2009ApJ...693.1392S} (the polarisation fraction ranges $1-15\%$ with an average of 4.2\,\%). The deep ($>15\,\mu$Jy) polarisation analysis at 1.4 GHz of the GOODS-N field shows a flattening of the polarised counts $\d\log N(>p)/\d\log (p)$ from $-1.5$ to $-0.6$ below 1\,mJy \citep{2014ApJ...785...45R}, which is ascribed to the star-forming population being typically less polarised than AGNs.

Converting from polarisation fractions to $f_{\rm pol}$ (the number of galaxies detected in both total intensity and polarisation) is not straightforward. One crucial aspect is the extent of the anti-correlation between total intensity and fractional polarisation found in many studies \citep[e.g.][]{2007ApJ...666..201T,2010MNRAS.402.2792S}. The most recent results report a flattening of this anti-correlation at fainter fluxes, which would limit $f_{\rm pol}$ to a few percent \citep[e.g.][]{2014ApJ...785...45R,2014ApJ...787...99S}.  

There are indications that the polarisation angle is strongly aligned with the galaxies' major axis, with intrinsic scatter lower than $\pm 15^\circ$ \citep{2012A&A...543A.127S,2009ApJ...693.1392S}. Reasonable physical considerations suggest that this scatter will be lower in galaxies which have stronger magnetic fields and hence higher levels of polarisation, giving a useful correlation between polarisation fraction of a sample and $\alpha_{\rm rms}$. Furthermore, even in cases where polarisation is not formally detected to a given significance in a blind polarisation survey, we will be attempting to measure polarisation of objects detected in continuum surveys. This continuum detection may be expected to give useful prior information on a polarisation measurement (e.g. the sky location of the galaxy). Forming a full posterior probability $P(\alpha)$ for the polarisation angle should then be possible and could then be propogated to the shear estimator regardless of the headline detection significance.

\section{Non-Linear Scales}\label{sec:sys_nl}
The final major source of systematic errors that we investigate in this work is due to an incorrect treatment of the non-linear r\'egime of perturbations. Recently, several analyses have tried to asses the impact of such systematics on state-of-the-art cosmic shear surveys such as the CFHTLenS \citep[see e.g.][]{Kitching:2014dtq,Ade:2015rim}. Since the non-linear evolution of density fluctuations affects mostly the smaller (angular) scales, the simplest way to avoid having to deal with non-linear perturbations is therefore to limit the analysis to large cosmic scales. It is also well-known that the physical scales at which linear theory does not hold any longer, $k_{\rm nl}$, is a redshift-dependent quantity, which monotonically increases with redshift. That is to say, the deeper the survey the larger $k_{\rm nl}$, i.e.\ the more the number of modes available in the linear r\'egime. Moreover, another effect of non-linearities is to increase the covariance matrix, in particular its non-Gaussian and super-sample variance terms. This clearly represents an additional reason for which a correct treatment of non-linear scales is imperative not to assess wrongly the constraining power of a future survey.

\begin{figure}
\includegraphics[width=\columnwidth]{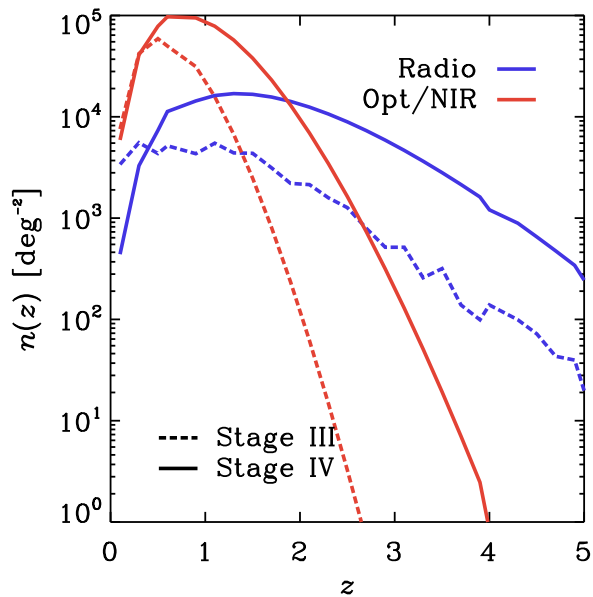}
\caption{Redshift distribution of sources per square degree.}\label{fig:dNdz}
\end{figure}

In this respect, Fig.~\ref{fig:dNdz} shows the redshift distributions of radio (blue curves) and optical/near-IR (red curves) sources for Stage III and Stage IV DETF cosmic shear surveys (dashed and solid lines, respectively). Clearly, although radio surveys will in general detect fewer sources than their optical counterparts, the former exhibit a significant high-redshift tail. For example, SKA1 goes as deep as a \euclid\ survey does, whilst the full SKA will observe non-negligible number densities for sources at $z>3$. For a start, this represents a major complementarity between cosmic shear experiments in the two bands. Moreover, it is a clear indication in favour of radio-optical cross-correlation.

In this Section, we are interested in quantifying how such a longer lever arm possessed by radio experiments affects the robustness of cosmological analyses.

\subsection{Amount of Information from Linear Scales}
\begin{figure*}
\includegraphics[width=\textwidth]{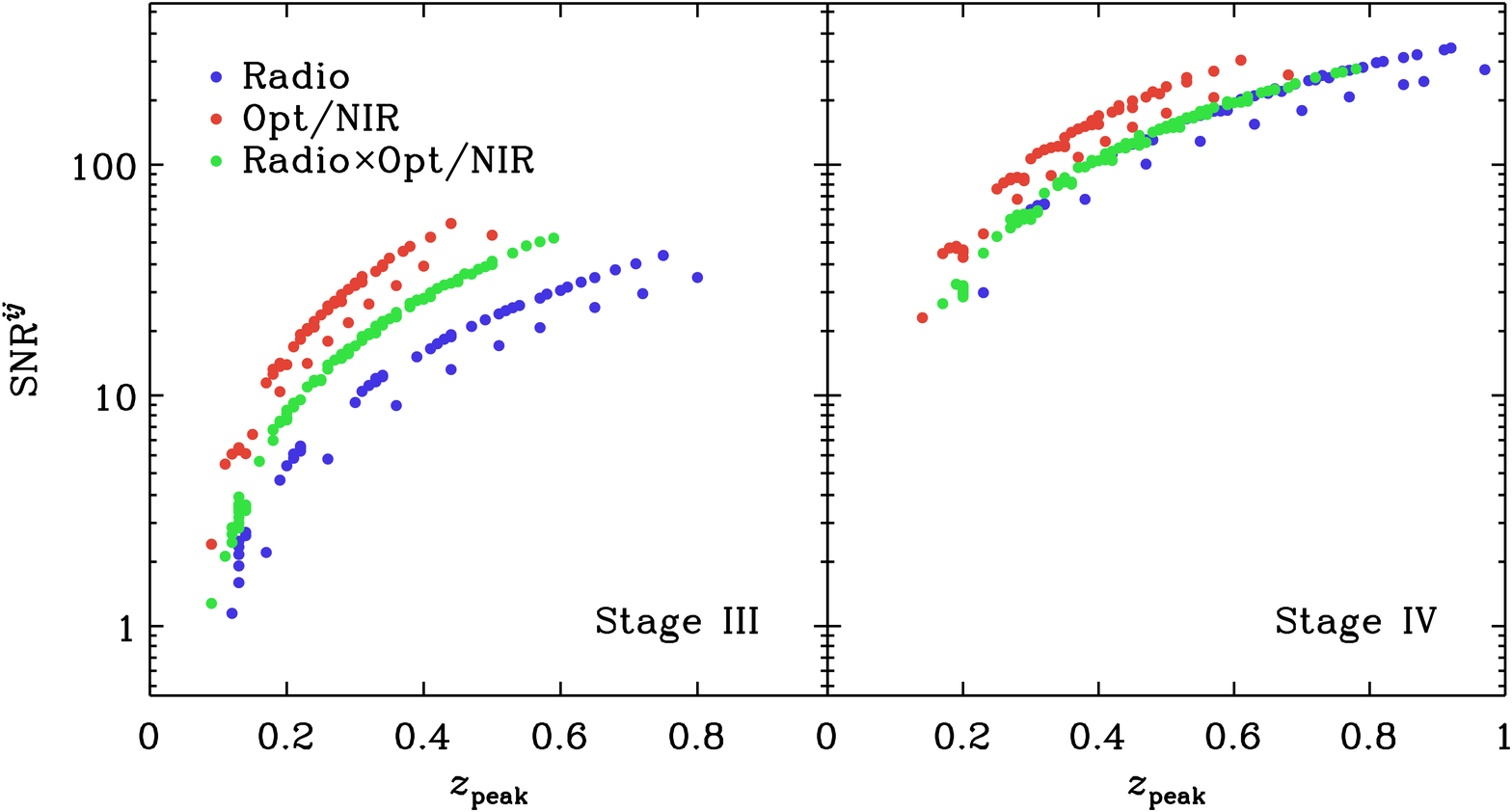}
\caption{SNR as a function of peak redshift.}\label{fig:snr_zp}
\end{figure*}

A way to visualise this \citep[cfr][Fig.~3]{Heymans:2013fya} is to compute, for each bin pair, the peak redshift $z_{\rm peak}$ corresponding to the comoving distance at which the product of the weak lensing kernels for the two bins, $W^{X_i}(z)W^{Y_j}(z)$, peaks. Then, for each power spectrum we calculate the total signal-to-noise ratio of the forecast measurement as
\begin{equation}
\snr^{ij}=\sqrt{\sum_\ell\frac{\left(C^{X_iY_j}_\ell\right)^2}{\sigma^2\left(C^{X_iY_j}_\ell\right)}},\label{eq:snr}
\end{equation}
where, if the signal is $\mathbf s$, $\sigma^2\left(\mathbf s\right)$ is the variance on its measurement. Figure~\ref{fig:snr_zp} shows the \snr\ for each bin pair $X_i-Y_j$ as a function of the corresponding peak redshift for Stage III and Stage IV DETF experiments (left and right panel, respectively). It is straightforward to see that radio experiments (blue points) reach higher redshifts compared to their optical/near-IR counterparts---although the smaller source number density of SKA1 leads to overall smaller \snr s compared to DES. The highest-redshift bins of radio catalogues contain galaxies 30 to 50 per cent more distant than sources in their optical/near-IR counterparts. Green points show the results for the cross-correlation \textit{only} of radio and optical/near-IR experiments.

\begin{figure*}
\includegraphics[width=\textwidth]{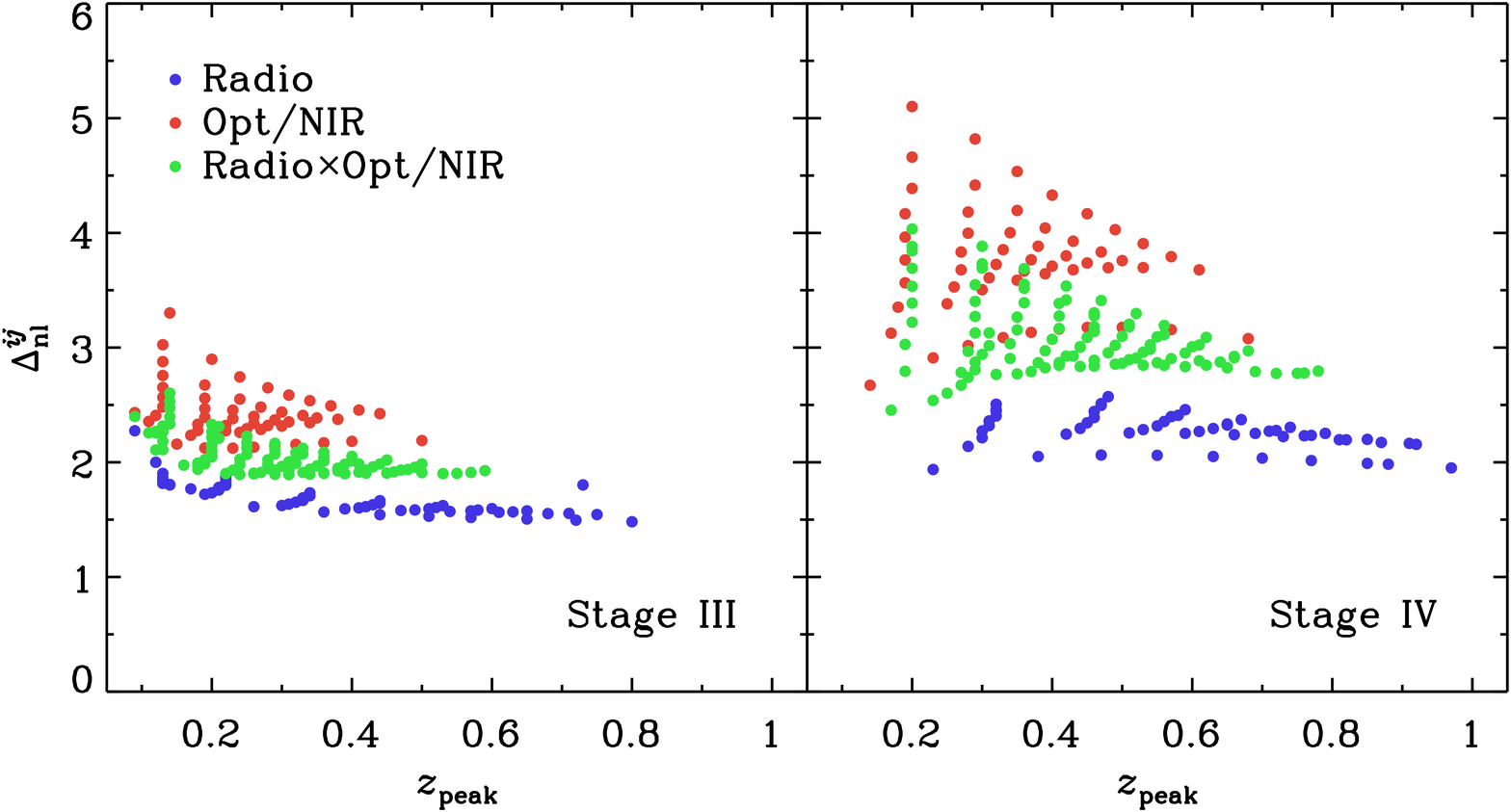}
\caption{Amount of SNR (squared) from non-linear scales with respect to linear theory.}\label{fig:flin_zp}
\end{figure*}

Then, to quantify better the impact of such high-redshift information for cosmological analyses, we also compute the same \snr\ of Eq.~\eqref{eq:snr} but considering only the linear matter power spectrum. In this case, we define a new metric
\begin{equation}
    \Delta^{ij}_{\rm nl}=\frac{\left(\snr^{ij}\right)^2}{\left(\snr^{ij}_{\rm lin}\right)^2},\label{eq:Delta_nl}
\end{equation}
where the subscript `lin' refers to the \snr\ of Eq.~\eqref{eq:snr} computed only using the angular cosmic shear power spectrum from linear theory.\footnote{Note that this is not the same as saying that we only consider information coming from linear scales.} The results are shown in Fig.~\ref{fig:flin_zp}. This gives us a proxy for the fraction of information coming from non-linear scales, which have to be treated more carefully because the poorly understood effect of baryons and, in general, the non-linear growth of structures demands some degree of \textit{ad hoc} modelling \citep{Semboloni:2012yh,Kitching:2014dtq,Fedeli:2014gja,Fedeli:2014hca}. What this plot tells us is that, in the fixed multipole range $20\leq\ell\leq3000$, the total SNR for the cosmic shear angular power spectrum in a given redshift bin pair is $\Delta^{ij}_{\rm nl}$ times what we would get if all the scales considered were linear. Consider for instance the highest redshift for DES and SKA1 (left panel, rightmost red and blue points): the information DES actually measures is more than twice the na\"ive linear-theory prediction, whereas for SKA1 this is $\sim1.5$ times higher. In other words, the DES measurement is contaminated by more than $100\%$ by information from non-linear scales, whilst SKA1 only by $\sim50\%$. We want to emphasise that the case considered regards the highest-redshift combination of bins, and it is clear from the spread of the points in the plot that for auto- and, especially, cross-correlations of lower-redshift bins the spread between optical/near-IR and radio surveys is even more pronounced. Moreover, the right panel of Fig.~\ref{fig:flin_zp} demonstrates that the high fidelity of SKA cosmic shear experiments to linear theory is even higher for Stage IV DETF experiments.

\subsection{Constraining Dark Energy at Pivot Redshift}\label{ssec:fom_flin}
Another way to assess how much cosmological parameter estimation from a survey will be affected by information encoded in linear/non-linear scales is to study the  experiment sensitivity to dark energy at the redshift at which the tightest constraints on its equation of state can be achieved. This is the so-called `pivot redshift', $z_p$ \citep{Albrecht:2009ct}. If we recast the dark energy equation of state as
\begin{align}
    w(a)&=w_0+w_a(1-a)\\
        &=w(a_p)+w_a(a_p-a),
\end{align}
where we remind the reader that the scale factor $a(z)=1/(1+z)$ and $a_p=a(z_p)$, then through simple algebraic relations we get
\begin{equation}
    z_p=-\frac{\left(\mathbfss F^{-1}\right)_{w_0,w_a}}{\left(\mathbfss F^{-1}\right)_{w_0,w_a}+\left(\mathbfss F^{-1}\right)_{w_a,w_a}}.\label{eq:zpivot}
\end{equation}
The forecast marginal error on $w(a_p)\equiv w_p$ then reads
\begin{equation}
    \sigma(w_p)=\sqrt{\left(\mathbfss F^{-1}\right)_{w_0,w_0}-\frac{\left[\left(\mathbfss F^{-1}\right)_{w_0,w_a}\right]^2}{\left(\mathbfss F^{-1}\right)_{w_a,w_a}}}.
\end{equation}

\begin{figure}
\includegraphics[width=\columnwidth]{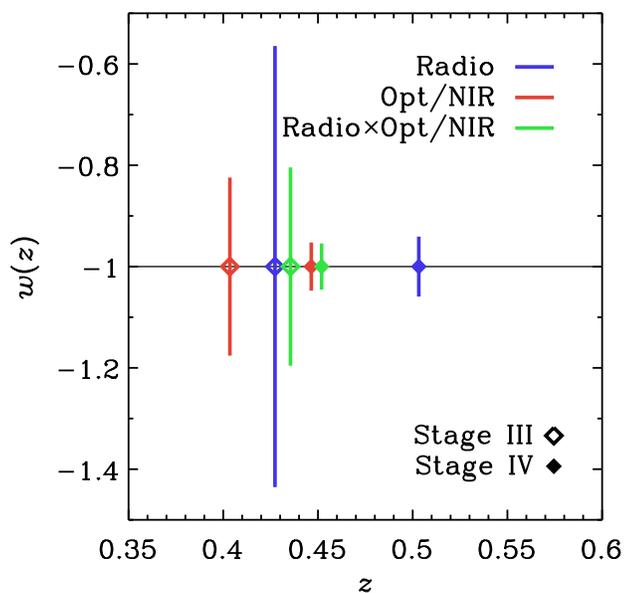}
\caption{Constraints on $w_p$ as a function of redshift.}
\label{fig:DEEoS_zp}
\end{figure}
Fig.~\ref{fig:DEEoS_zp} shows the forecast constraints on the equation-of-state parameter at the pivot redshift, $w_p=w(z_p)$, for the various experiments considered. Again, blue, red and green are for radio and optical/near-IR surveys and their cross-correlation. We can easily notice the tightness of constraints from Stage IV DETF experiments compared to that of Stage III surveys. On the other hand, it may seem unnatural that the pivot redshift from the cross-correlation of DES and SKA1 is larger than that of each survey alone. However, this is primarily due to different degeneracies among cosmological parameters in the various experiments. As different survey combinations probe the Universe's geometry and the growth of cosmic structures on different redshift ranges, their sensitivity to certain cosmological parameters also varies. To appreciate this effect, in Table~\ref{tab:zpivot} we compare the pivot redshift calculated according to Eq.~\eqref{eq:zpivot} with the full Fisher matrix, $\mathbfss F_{\alpha\beta}$ (left column), to that obtained using the Fisher matrix for $\{w_0,w_a\}$ only, $\mathbfss F_{w_0,w_a}$ (right column). Not being marginalised over all the other cosmological parameters, the latter is insensitive to degeneracies between dark energy and the other \lcdm\ parameters. Thus, we can now appreciate how, in principle, the deeper the median redshift of the experiment, the higher the redshift at which dark energy is most tightly constrained.
\begin{table}
\centering
\caption{\label{tab:zpivot}Comparison between $z_p$ from the full and the dark-energy only Fisher matrices.}
\begin{tabular}{lcc}
\hline
& $\mathbfss F_{\alpha\beta}$ & $\mathbfss F_{w_0,w_a}$\\
\hline
SKA1 & 0.43 & 0.31 \\
DES & 0.40 & 0.22 \\
SKA1$\times$DES & 0.44 & 0.25 \\
\hline
SKA2 & 0.50 & 0.37 \\
\euclid\ & 0.45 & 0.28 \\
SKA2$\times$\euclid\ & 0.45 & 0.32 \\
\hline
\end{tabular}
\end{table}

\section{Discussion and Conclusions}\label{sec:conclusions}
This paper is the third of a series on weak lensing cosmology with the SKA (Square Kilometre Array). In Paper I we have compared the forecast constraining power of Stage III and IV DETF cosmic shear surveys in the radio band with their equivalent at the optical/near-IR wavelengths. Then, in Paper II we have constructed a realistic pipeline for simulating SKA cosmological weak lensing surveys. Here, we focus on real-world effects such as contamination by various systematics and ways to avoid them---in particular, by exploiting radio information and synergies between radio and optical/near-IR measurements.

It is worth noting that the multi-wavelength cosmic shear experiments will have a major advantage in mitigating systematic effects, thanks to the fact that we can analyse auto- and cross-correlations between radio and optical/near-IR surveys separately. This can be employed to test for new, unaccounted for systematics. Let us consider the case where, on the one hand, the best-fit values of certain cosmological parameters, as reconstructed from either radio or optical/near-IR experiments alone, are many $\sigma$'s different from each other. On the other hand, the cross-correlation of the two experiments gives results in better agreement, say, with CMB measurements or other data sets. Then, this could imply that either or both of the two auto-correlation surveys suffer from some kind of unknown systematic effect, which in turn does not affect the cross-correlation.

To cover the widest possible range of systematic effects, we scrutinise the impact on cosmological parameter estimation of three different types of systematic errors: $(i)$ experimental systematics; $(ii)$ intrinsic alignments; and $(iii)$ modelling of non-linear scales. In the following, we draw the major conclusions for each of these sub-topics.
\begin{itemize}
    \item[$i)$] Experimental systematics will mostly affect the measured shear, $\gamma^{\rm obs}=\gamma+\gamma^{\rm sys}$, and their effects can be decomposed into an additive and a multiplicative term such that $\gamma^{\rm sys}=\gamma A_{\rm add}+\gamma^{\rm mul}\gamma$. The former may be regarded as a bias, whilst the latter as a calibration error. In this respect, the major advantage of pursuing weak lensing measurements in more than one band is represented by the fact that we can thus correlate radio and optical/near-IR shear catalogues to obtain a cross-correlation cosmic shear power spectrum. Since systematic errors for two completely different experimental setups will not correlate, the cross-correlation is expected to be free of additive systematics. This major result is presented in Fig.~\ref{fig:sys_add1}. The case for multiplicative systematics, though, is more subtle. This is because an overall calibration error will not cancel when correlating radio and optical/near-IR data. However, this too can be strongly alleviated by means of self-calibration, where we put together all the information from auto-correlations of radio and from optical/near-IR catalogues as well their cross-correlation, as shown in Fig.~\ref{fig:sys_mul}. This is a major result of this paper.
    \item[$ii)$] A na\"ive approach to cosmic shear assumes that the intrinsic distribution of galaxy ellipticities is random across the sky. However, this is known not to be true, as galaxies form within the large-scale cosmic structure and thus experience the same gravitational potential. As a result, physically close galaxies will be preferentially aligned with each other, whilst foreground galaxies shaped by a particular gravitational potential are expected to anti-correlate with background galaxies lensed by the same potential. A possible way to tackle this problem without losing much of a survey's constraining power is self-calibration, where the clustering information on galaxies' three-dimensional position is employed to constrain a set of IA nuisance parameters \citep[e.g.][]{Kirk:2011aw}. Following \citet{Brown:2010rr}, we here investigate a thoroughly different, though complementary, approach. It relies on the polarisation information additionally available from radio weak lensing measurements, and on the fact that a galaxy's polarised emission is a proxy for the galaxy's intrinsic orientation but is unaffected by gravitational lensing. Figure~\ref{fig:ellipses_IAs} illustrates that, depending on the fraction of sources for which polarisation information is available and on the intrinsic scatter of measurements of galaxy position angles, radio cosmic shear and its cross-correlation with optical/near-IR data is capable of recovering much of the cosmological information even when a large number of IA nuisance parameters is included not to bias the parameter estimation.
    \item[$iii)$] Cosmological analyses with the first generations of weak lensing surveys have shown that a proper understanding of non-linear scales is of utmost importance for cosmic shear to compete with other cosmological probes like the CMB or galaxy clustering. Concerning this, lensing in the radio band has the added value of the high-redshift tail of the radio source redshift distributions (see Fig.~\ref{fig:dNdz}). This implies that the range of linear scales available to radio cosmic shear surveys is larger than that of optical/near-IR experiments, as shown in Figs~\ref{fig:snr_zp}-\ref{fig:flin_zp}. Therefore, radio weak lensing appears to be less prone to systematic errors due to an incorrect modelling of non-linear scales, and can be used as a cross-check for optical/near-IR surveys.
\end{itemize}

\section*{Acknowledgments}
SC wishes to thank Alan Heavens for useful discussions. SC, IH, AB and MLB are supported by ERC Starting Grant No. 280127. MLB is an STFC Advanced/Halliday fellow.

\bibliographystyle{mn2e_plus_arxiv}
\bibliography{main_bib}

\appendix
\section{Bias on Parameter Estimation}\label{sec:Fisher_bias}
To compute the bias expected on the best-fit value of a cosmological parameter $\vartheta_\alpha$ due to the neglect of a systematic effect, we follow a Fisher matrix approach based on the concept of `nested models' \citep[see e.g.][]{Heavens:2007ka,Camera:2010wm}. Two models are nested when the parameter space of the former (or `simpler' model) is contained within that of the latter. A straightforward example of this is \lcdm\ and dark energy: as presented at the end of Sec.~\ref{sec:fisher}, the \lcdm\ parameter space is a hypersurface of that of dark energy. Namely, the dark energy equation-of-state parameters effectively represent two additional directions in the \lcdm\ parameter space.

Such a framework for nested models can be easily recast when dealing with systematic effects. In this case, the additional term in the observable due to systematics---the additive and multiplicative systematics power spectra or the IA GI and II terms---can be parameterised by an overall amplitude $f_{\rm sys}$. This is a fudge factor, either equal to 1 (if the systematic effect is actually present) or 0. Then, the parameter vector of our model becomes $\{\vartheta_\alpha\}\cup\{f_{\rm sys}^i\}$, where $\{\vartheta_\alpha\}$ is the set of the cosmological parameters of interest, and $\{f_{\rm sys}^i\}$ is the set of all the possible systematic effect fudge factors.

Then, if the correct underlying model does contain certain systematics, in the incorrect model where we neglect them, the maximum of the expected likelihood will not, in general, be at the correct parameter values \citep[see][Fig.~1]{Heavens:2007ka}. The parameters of the incorrect model will have to shift their values to compensate the fact that $\{f_{\rm sys}^i\}$ are being kept fixed at the incorrect fiducial value $\{f_{\rm sys}^i=0\}$. We can compute these shifts, i.e.\ the biases in the cosmological parameter best-fit values, according to
\begin{equation}
b(\vartheta_\alpha)=\left(\mathbfss H^{-1}\right)_{\alpha\beta}\mathbfss G_{\beta i},\label{eq:bias}
\end{equation}
where $\mathbfss G$ is a sub-matrix of the Fisher matrix for the full parameter set $\{\vartheta_\alpha\}\cup\{f_{\rm sys}^i\}$. (Note that summation over equal indices is assumed here.) Now, it is worth spending a few words on $\mathbfss H$, as in the literature different approaches have been followed. They can basically be connected to two cases: whether $(i)$ $\mathbfss H$ is the Fisher matrix for the cosmological parameters only, or $(ii)$ it is a subset of the Fisher matrix of the full parameter set. To understand this, we shall consider a simple two-dimensional scenario of a Gaussian likelihood $L=-\chi^2/2$. The chi-square reads
\begin{equation}
\chi^2(\mathbf x)=(\mathbf x-\bmu)\mathbfss F_\textrm{2D}(\mathbf x-\bmu)^T,
\end{equation}
with $\mathbf x=(x,y)$ and $\mathbfss F_\textrm{2D}=\boldsymbol\Sigma_\textrm{2D}^{-1}$. Here,
\begin{equation}
\boldsymbol\Sigma_\textrm{2D}=\left(\begin{array}{cc}
    \sigma_x^2 & \varrho\sigma_x\sigma_y \\
    \varrho\sigma_x\sigma_y & \sigma_y^2
\end{array}\right)
\end{equation}
is the covariance matrix. The minimum of the chi-square is obviously $\bmu=(\mu_x,\mu_y)$. However, if we now disregard one of the parameter axes, we cut the chi-square surface along, say, $y=0$. In this case, it is easy to compute the minimum of $\chi^2(x,y=0)$, which is not $\mu_x$ but 
\begin{equation}
\mu_x-\frac{\varrho\sigma_x}{\sigma_y}\mu_y.
\end{equation}
The difference between the true minimum along the $x$-axis and the one along the $y=0$ surface is what we call the bias in the reconstruction of $\mu_x$. Now, if we take Eq.~\eqref{eq:bias} with $\mathbfss H=(\mathbfss F_\textrm{2D})_{1,1}$ and $\mathbfss G=(\mathbfss F_\textrm{2D})_{1,2}$, we find exactly the result above---where, in that case, $\mu_y=f_{\rm sys}\equiv1$. This means that $\mathbfss H$ is a subset of the full, cosmological$+$systematics Fisher matrix. Instead, the use of $\mathbfss H$ as the Fisher matrix for the cosmological parameters only would correspond to $\mathbfss H=(\mathbfss F_\textrm{1D})_{1,1}\equiv1/\sigma_x^2$.

\section{Stability of Numerical Derivation for Fisher Matrices}\label{sec:Fisher_stability}
One of the most important factors for the reliability of Fisher matrices is the stability of the numerical derivatives of the observables with respect to the cosmological parameters, ${\partial \mathbfss{C}^{XY}_\ell}/{\partial \vartheta_\alpha}$. For the sake of clarity, let us simplify the notation and consider a single observable $f$, function of a single parameter $p$. A common procedure in this regard is to use the so-called five point stencil, which involves the computations of $f(p)$ at five values of $p$, evenly spaced in a neighbourhood of the fiducial value, $\overline p$. However, the step size $\delta p$ by which one samples the $p$-line is critical, because if such step size is too large, the incremental ratio
\begin{equation}
\frac{-f(\overline p+2\delta p)+8f(\overline p+\delta p)-8f(\overline p-\delta p)+f(\overline p-2\delta p)}{12\delta p}\label{eq:4pointDerivative}
\end{equation}
is not a good proxy of the true derivative
\begin{equation}
\left.\frac{\d f}{\d p}\right|_{p=\overline p}=\lim_{\delta p\to0}\frac{f(\overline p+\delta p)-f(\overline p-\delta p)}{\delta p}.
\end{equation}
On the other hand, if $\delta p$ is too small, numerical instabilities render Eq.~\eqref{eq:4pointDerivative} unreliable. Such instabilities are due to the fact that one subtracts two numbers, $f(\overline p+\delta p)$ and $f(\overline p-\delta p)$, that are almost equal. Therefore, it is clear that the choice of $\delta p$ is of prime importance.

The scenarios investigated in the present paper involve a large number of observables and several cosmological parameters. More specifically, we have to deal with auto- and cross-correlations of 10 redshift bins for the SKA (both for Phase 1 and Phase 2) and 10 redshift bins for DES/\euclid, calculated for $\sim150$ multipoles $\ell$ and for a set of 7 cosmological parameters $\{\vartheta_\alpha\}$ plus up to 40 additional nuisance parameters. As a result, a case-by-case check for all the ${\partial \mathbfss{C}^{XY}_\ell}/{\partial \vartheta_\alpha}$ is highly time consuming and may result in a bin- and $\ell$-dependent optimal step size for the steps, $\delta\vartheta_\alpha$, depending on how much the observables is sensitive to a specific parameter, in a specific redshift bin and at a given angular scale.

To overcome this issue and check systematically the numerical derivative stability, we have therefore implemented an alternative pipeline, which works as follows:
\begin{itemize}
\item[$i)$] For each combination of experiments and bin pair, $X_i-Y_j$, multipole, $\ell$, and cosmological parameter, $\vartheta_\alpha$, we sample the $\vartheta_\alpha$-line in 15 points around $\overline{\vartheta_\alpha}$ (this included). More precisely, we take $\delta\vartheta_\alpha=0$, $\pm0.625\%$, $\pm1.25\%$, $\pm1.875\%$, $\pm2.5\%$, $\pm3.75\%$, $\pm5\%$ and $\pm10\%$.
\item[\textit{ii.a)}] In the hypothesis that the neighbourhood is small enough around $\overline{\vartheta}$, all the $C^{X_iY_j}_\ell(\vartheta_\alpha)$ thus obtained should lie on a straight line. We test this ansatz by testing if the spread between the linearly fitted $[C^{X_iY_j}_\ell(\vartheta_\alpha)]^{\rm fit}$ and the true values $[C^{X_iY_j}_\ell(\vartheta_\alpha)]^{\rm true}$ is less than $1\%$.
\item[\textit{ii.b)}] If this requirement is not met, we zoom in on the sampled $\vartheta_\alpha$-range by cutting out a few values on the edges, until we reach the requested accuracy.
\item[$iii)$] For each given combination $\{X_i,\,Y_j,\,\ell,\,\vartheta_\alpha\}$, the numerical derivative ${\partial C^{X_iY_j}_\ell}/{\partial \vartheta_\alpha}$ is therefore the slope of the linear interpolation.
\end{itemize}

Lastly, we have made a further final check to ensure that not only the numerical derivatives are reliable, but also the Fisher matrices thus calculated are stable with respect to our procedure. To do so, we have recomputed the Fisher matrices by randomly varying each single ${\partial C^{X_iY_j}_\ell}/{\partial \vartheta_\alpha}$---i.e.\ the various slopes of the linear fits---by $\pm1\sigma$ estimated errors on the fit slope. Then, we have compared the forecast marginal errors on the cosmological parameters, $\sigma(\vartheta_\alpha)$, in the two cases and we have found that the difference is negligible.

\section{Validation of the Fisher Procedure and Comparison with Paper I}\label{sec:Fisher_comparison}
In Paper I, we made a first comparison between MCMCs and Fishers matrices. We considered a number of simplifying factors, running MCMCs in each case and comparing the marginal error on parameters for each run between the Fisher matrix and MCMC (see Paper I, Sec. 4.2 for additional details). Here, we proceed further by exploring the full parameter space. To have a better comparison with MCMCs, we here include broad Gaussian priors mimicking the flat priors used in the MCMC analysis. Specifically, these Gaussian priors are such that their 1$\sigma$ errors correspond to the edges of the top-hat function used for the flat priors. In Table~\ref{tab:Fisher-vs-MCMCs}, we present a comparison between forecast marginal errors obtained from MCMCs (from Paper I) and the Fisher matrices computed in the present work.
\begin{table*}
\centering
\caption{\label{tab:Fisher-vs-MCMCs}Forecast marginal errors from MCMCs (Paper I) and Fisher matrices (present work).}
\begin{tabular}{lccccccccc}
\hline 
& \multicolumn{2}{c}{$\sigma(\om)/\om$} & \multicolumn{2}{c}{$\sigma(\sigma_{8})/\sigma_8$} & & \multicolumn{2}{c}{$\sigma(\w)$} & \multicolumn{2}{c}{$\sigma(\wa)$}\\
\cline{2-5}\cline{7-10}
& MCMC & Fisher & MCMC & Fisher & & MCMC & Fisher & MCMC & Fisher\\
SKA1 &                          0.083 & 0.082 & 0.040 & 0.041 && 0.52 & 0.54 & 1.6 & 1.6\\
SKA1 + \planck &                0.084 & 0.080 & 0.040 & 0.040 && 0.28 & 0.28 & 0.43 & 0.43\\
DES &                           0.056 & 0.057 & 0.032 & 0.033 && 0.43 & 0.46 & 1.4 & 1.5\\
DES + \planck &                 0.058 & 0.057 & 0.033 & 0.033 && 0.22 & 0.22 & 0.33 & 0.34\\
SKA1$\times$DES &               0.046 & 0.053 & 0.024 & 0.030 && 0.45 & 0.45 & 1.3 & 1.4\\
SKA1$\times$DES + \planck &     0.046 & 0.053 & 0.024 & 0.029 && 0.23 & 0.23 & 0.36 & 0.36\\
\hline
SKA2 &                          0.010 & 0.011 & 0.0046 & 0.0049 && 0.14 & 0.17 & 0.42 & 0.48\\
SKA2 + \planck &                0.010 & 0.011 & 0.0047 & 0.0049 && 0.086 & 0.11 & 0.15 & 0.18\\
\euclid &                       0.011 & 0.012 & 0.0058 & 0.059 && 0.13 & 0.14 & 0.38 & 0.44\\
\euclid + \planck &             0.012 & 0.012 & 0.0059 & 0.058 && 0.095 & 0.085 & 0.16 & 0.15\\
SKA2$\times$\euclid &           0.013 & 0.010 & 0.0064 & 0.0049 && 0.15 & 0.13 & 0.43 & 0.39\\
SKA2$\times$\euclid + \planck & 0.013 & 0.010 & 0.0064 & 0.0048 && 0.10 & 0.082 & 0.17 & 0.14\\
\hline
\end{tabular}
\end{table*}

\begin{figure*}
\includegraphics[width=\textwidth]{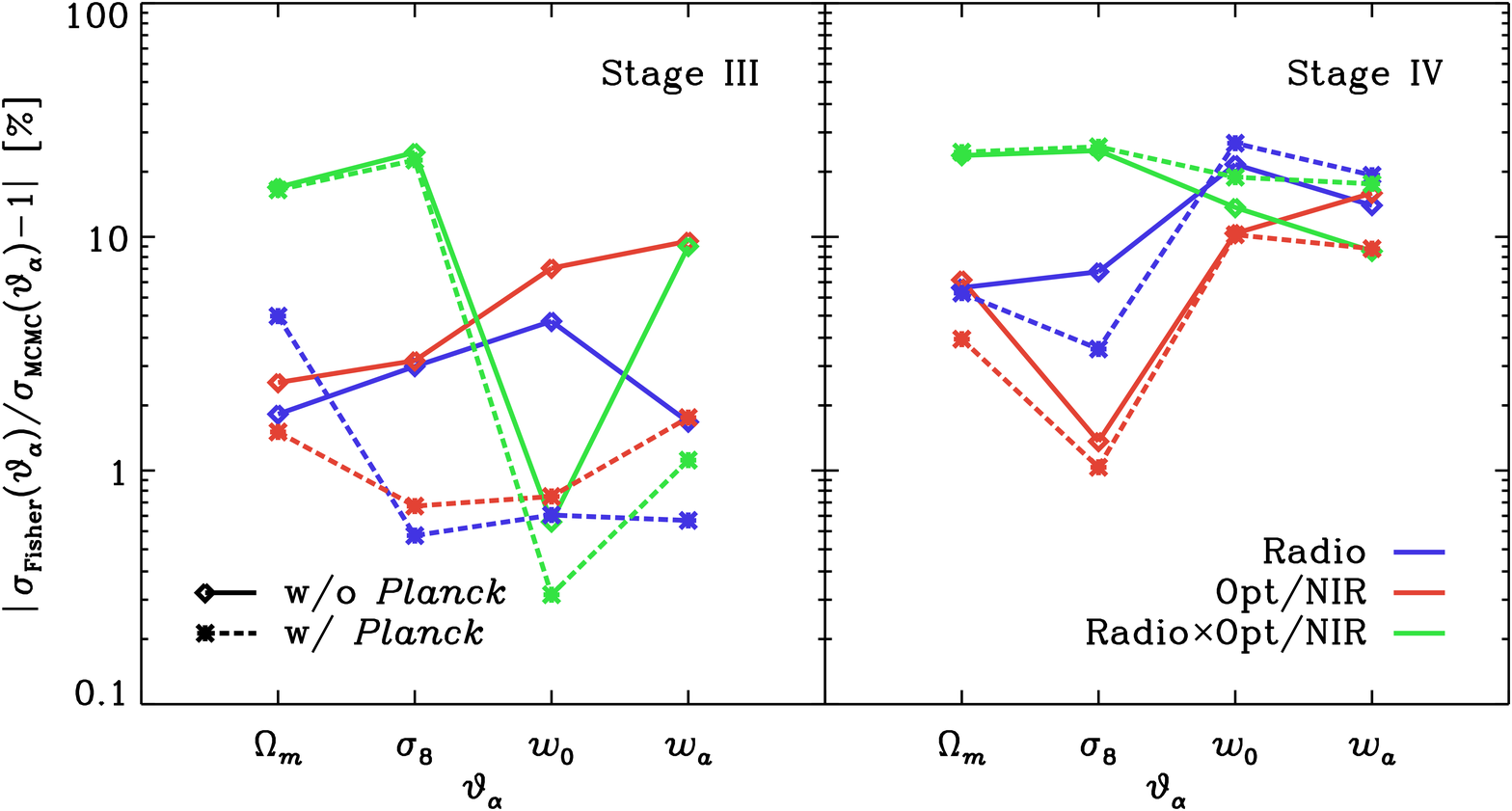}
\caption{Percentual relative difference between between forecasts from MCMCs (Paper I) Fisher matrices (present work).}
\label{fig:FMvsMCMC}
\end{figure*}

A visual presentation of the same results is given in Fig.~\ref{fig:FMvsMCMC}, where the percentual relative difference between 1$\sigma$ Fisher matrix forecast marginal errors and 68\% MCMC estimated confidence intervals on cosmological parameters is shown for all the parameters present in Table~\ref{tab:Fisher-vs-MCMCs}. Left and right panels are respectively for Stage III and Stage IV DETF cosmic shear experiments, with results from radio spectra in blue, from optical/near-IR ones in red and their cross-spectra depicted in green. Solid and dashed lines are only eye-guides to differentiate between the inclusion or not of priors from \planck. The agreement is extremely good, with a scatter much smaller than $10\%$ for most of the configurations, in particular for Stage III surveys \citep[cfr][]{Wolz:2012sr}. We also find some discrepancies between the two methods, for instance when non-Gaussian contours are involved---as for the well-known $\om$-$\sigma_8$ degeneracy---or when priors become particularly important---as for dark energy parameters.

\bsp

\label{lastpage}

\end{document}